\documentclass[final]{IEEEtran}

\usepackage{amsmath,amssymb,bm}
\usepackage{graphicx}
\usepackage{cite,url}

\newlength\figwidth
\begin{document}

\title{On the Design of Perceptual MPEG-Video Encryption Algorithms%
\thanks{Copyright (c) 2006 IEEE. Personal use of this material is
permitted. However, permission to use this material for any other
purposes must be obtained from the IEEE by sending an email to
\texttt{pubs-permissions@ieee.org}.}
\thanks{This research was partially supported by the City University
of Hong Kong SRG grant 7001702, by The Hong Kong Polytechnic
University's Postdoctoral Fellowships Scheme under grant no. G-YX63,
, by the Research Grant Council of Hong Kong under grant no. PolyU
5232/06E, and by the US NSF grants ANI-0219110 and RIS-0292890.}}
\author{Shujun Li\thanks{Shujun Li and Kwok-Tung Lo are with the
Department of Electronic and Information Engineering, The Hong Kong
Polytechnic University, Hung Hom, Kowloon, Hong Kong SAR, China.},
Guanrong Chen,~\IEEEmembership{Fellow, IEEE}\thanks{Guanrong Chen is
with the Department of Electronic Engineering, City University of
Hong Kong, 83 Tat Chee Avenue, Kowloon Tong, Hong Kong SAR, China.},
Albert Cheung,~\IEEEmembership{Member, IEEE}\thanks{Albert Cheung is
with the Department of Building and Construction and Shenzhen
Applied R\&D Centres, City University of Hong Kong, Kowloon Tong,
Hong Kong SAR, China.}, Bharat Bhargava,~\IEEEmembership{Fellow,
IEEE}\thanks{Bharat Bhargava is with the Department of Computer
Sciences, Purdue University, 250 N. University Street, West
Lafayette, IN 47907-2066, USA.} and Kwok-Tung
Lo,~\IEEEmembership{Member, IEEE}}

\markboth{IEEE TRANSACTIONS ON CIRCUITS AND SYSTEMS FOR VIDEO
TECHNOLOGY, VOL. 17, NO. 2, PAGES 214-223, FEBRUARY 2007}{Shujun Li
\MakeLowercase{\textit{et al.}}: Perceptual MPEG Encryption}

\maketitle

\begin{abstract}
In this paper, some existing perceptual encryption algorithms of
MPEG videos are reviewed and some problems, especially security
defects of two recently proposed MPEG-video perceptual encryption
schemes, are pointed out. Then, a simpler and more effective design
is suggested, which selectively encrypts fixed-length codewords
(FLC) in MPEG-video bitstreams under the control of three
perceptibility factors. The proposed design is actually an
encryption configuration that can work with any stream cipher or
block cipher. Compared with the previously-proposed schemes, the new
design provides more useful features, such as strict
size-preservation, on-the-fly encryption and multiple
perceptibility, which make it possible to support more applications
with different requirements. In addition, four different measures
are suggested to provide better security against
known/chosen-plaintext attacks.
\end{abstract}


\section{Introduction}

The wide use of digital images and videos in various applications
brings serious attention to the security and privacy issues today.
Many different encryption algorithms have been proposed in recent
years as possible solutions to the protection of digital images and
videos, among which MPEG videos attract most attention due to its
prominent prevalence in consumer electronic markets
\cite{Zeng:MultimediaSecurity:Book2006,
Ahl:ImageVideoEncryption:Book2005,
Furht:MultimediaSecurity:Book2005,
Furht:ImageVideoEncryption:Handbook2004,
Li:ChaosImageVideoEncryption:Handbook2004}.

\begin{figure}
\centering
\includegraphics[width=\columnwidth]{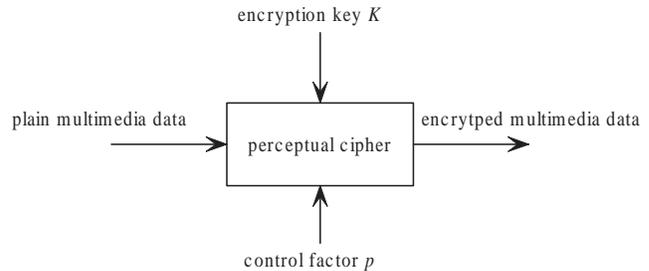}
\caption{A diagrammatic view of the perceptual
encryption.}\label{figure:PE}
\end{figure}

In many applications, such as pay-per-view videos, pay-TV and video
on demand (VoD), the following feature called ``perceptual
encryption" is useful. This feature requires that the quality of
aural/visual data is only \textit{partially} degraded by encryption,
i.e., the encrypted multimedia data are still partially perceptible
after encryption. Such perceptibility makes it possible for
potential users to listen/view low-quality versions of the
multimedia products before buying them. It is desirable that the
aural/visual quality degradation can be continuously controlled by a
factor $p$, which generally denotes a percentage corresponding to
the encryption strength. Figure~\ref{figure:PE} shows a diagrammatic
view of perceptual encryption. The encryption key is kept secret
(not needed when public-key ciphers are used) but the control factor
$p$ can be published.

Regarding the visual quality degradation of the encrypted videos,
the following points should be remarked: 1) since there does not
exist a well-accepted objective measure of visual quality of digital
images and videos, the control factor is generally chosen to
represent a rough measure of the degradation; 2) the visual quality
degradations of different frames may be different, so the control
factor works only in an average sense for all videos; 3) the control
factor is generally selected to facilitate the implementation of the
encryption scheme, which may not have a linear relationship with the
visual quality degradation (but a larger value always means a
stronger degradation); 4) when the control factor $p=1$, the
strongest visual quality degradation of the specific algorithm
(i.e., of the target application) is reached, but it may not be the
strongest degradation that all algorithms can produce (i.e., all
visual information of the video is completely concealed).

In recent years, some perceptual encryption schemes have been
proposed for G.729 speech
\cite{Servetti:PerceptualSpeech:ICASSP2002,
Servetti:PerceptualSpeech:IEEETSAP2002}, MP3 music
\cite{Torrubia:PerceptualMP3:IEEETCE2002}, JPEG images
\cite{Belgian:SelectiveImageEncryption:ACIVS2002,
Torrubia:PerceptualJPEG:ICCE2003}, wavelet-compressed (such as
JPEG2000) images and videos
\cite{Lian:PerceptualCryptography:ICME2004,
Lian:PerceptualCryptography:CIT2004,
Lian:PerceptualCryptography:ISIMP04} and MPEG videos
\cite{Dittmann:EnablingMPEG:LNCS97, Yann:WaterScrambling:ACMMM2002,
Turk:MPEG2Scrambling:IEEETCE2002,
Chinese:MPEG2Scrambling:IEEETCE2003}, respectively. The selective
encryption algorithms proposed in
\cite{Pommer&Uhl:WPSelectiveEncryption:SPS2002,
Pommer&Uhl:SelectiveWaveletEncryption:MS2003,
Pommer:SelectiveWaveletEncryption:Thesis2003} can be considered as
special cases of the perceptual encryption for images compressed
with wavelet packet decomposition. In some research papers, a
different term, ``transparent encryption", is used instead of
``perceptual encryption" \cite{Turk:MPEG2Scrambling:IEEETCE2002,
Chinese:MPEG2Scrambling:IEEETCE2003}, emphasizing the fact that the
encrypted multimedia data are \textit{transparent} to all
standard-compliant decoders. However, transparency is actually an
equivalent of another feature called ``format-compliance" (or
``syntax-awareness") \cite{Zeng:VideoScrambling:MMSP2001,
Zeng:VideoScrambling:IEEETCASVT2002}, which does not mean that some
partial perceptible information in plaintexts still remains in
ciphertexts. In other words, a perceptual cipher must be a
transparent cipher, but a transparent cipher may not be a perceptual
cipher \cite{Li:ChaosImageVideoEncryption:Handbook2004}. Generally,
perceptual encryption is realized by selective encryption algorithms
with the format-compliant feature. This paper chooses to use the
name of ``perceptual encryption" for such a useful feature of
multimedia encryption algorithms. More precisely, this paper focuses
on the perceptual encryption of MPEG videos. After identifying some
problems of the existing perceptual encryption schemes, a more
effective design of perceptual MPEG-video encryption will be
proposed.

The rest of this paper is organized as follows. The next section
will provide a brief survey of related work and point out some
problems, especially problems existing in two recently-proposed
perceptual encryption algorithms
\cite{Turk:MPEG2Scrambling:IEEETCE2002,
Chinese:MPEG2Scrambling:IEEETCE2003}. In Section
\ref{section:NewDesign}, the video encryption algorithm (VEA)
proposed in \cite{Shi:MPEGEncryption:MMTA2004} is generalized to
realize a new perceptual encryption design for MPEG videos, called
the perceptual VEA (PVEA). Experimental study is presented in Sec.
\ref{section:Experiments}, to show the encryption performance of
PVEA. The last section presents the conclusion.

\section{Related Work and Existing Problems}

\subsection{Scalability-based perceptual encryption}

Owing to the scalability provided in MPEG-2/4 standards
\cite{MPEG2-ISOStandard, MPEG4-ISOStandard}, it is natural to
realize perceptual encryption by encrypting the enhancement
layer(s) of an MPEG video (but leaving the base layer unencrypted)
\cite{Dittmann:EnablingMPEG:LNCS97}. However, since not all MPEG
videos are encoded with multiple layers, this scheme is quite
limited in practice. More general designs should be developed to
support videos that are compliant to the MPEG standards.

\subsection{Perceptual encryption for JPEG images}

Due to the similarity between the encoding of JPEG images
\cite{JPEG-ISOStandard} and the frame-encoding of MPEG videos
\cite{MPEG1-ISOStandard, MPEG2-ISOStandard, MPEG4-ISOStandard},
the ideas of perceptual encryption for JPEG images can be easily
extended to MPEG videos.

In \cite{Belgian:SelectiveImageEncryption:ACIVS2002}, two
techniques of perceptual encryption were studied: encrypting
selective bit-planes of uncompressed gray-scale images, and
encrypting selective high-frequency AC coefficients of JPEG
images, with a block cipher such as DES, triple-DES or IDEA
\cite{Schneier:AppliedCryptography96}. The continuous control of
the visual quality degradation was not discussed, however.

In \cite{Torrubia:PerceptualJPEG:ICCE2003}, the perceptual
encryption of JPEG images is realized by encrypting VLCs
(variable-length codewords) of partial AC coefficients in a ZoE
(zone of encryption) to be other VLCs in the Huffman table. The
visual quality degradation is controlled via an encryption
probability, $p/100\in[0,1]$, where $p\in\{0,\cdots,100\}$. This
encryption idea is similar to the video encryption algorithm
proposed in \cite{Zeng:VideoScrambling:IEEETCASVT2002}. The main
problem with encrypting VLCs is that the size of the encrypted
image/video will be increased since the Huffman entropy
compression is actually discarded in this algorithm.

\subsection{Perceptual encryption for wavelet-compressed images and
videos}

In \cite{Lian:PerceptualCryptography:ICME2004,
Lian:PerceptualCryptography:ISIMP04,
Lian:PerceptualCryptography:CIT2004}, several perceptual
encryption schemes for wavelet-compressed images and videos were
proposed. Under the control of a percentage ratio $q$, sign bit
scrambling and secret permutations of wavelet
coefficients/blocks/bit-planes are combined to realize perceptual
encryption. The problem with these perceptual encryption schemes
is that the secret permutations are not sufficiently secure
against known/chosen-plaintext attacks
\cite{Jan-Tseng:SCAN:IPL1996, Yu-Chang:SCAN:PRL2002,
Zhao:PositionPermute:ZJUS2004, Li:AttackingPOMC2004}: by comparing
the absolute values of a number of plaintexts and ciphertexts, one
can reconstruct the secret permutations. Once the secret
permutations are removed, the encryption performance will be
significantly compromised.

\subsection{Perceptual encryption of motion vectors in MPEG-videos}

In \cite{Yann:WaterScrambling:ACMMM2002}, motion vectors are
scrambled to realize perceptual encryption of MPEG-2 videos. Since
I-frames do not depend on motion vectors, such a perceptual
encryption algorithm can only blur the motions of MPEG videos. It
cannot provide enough degradation of the visual quality of the MPEG
videos for encryption (see Fig.~\ref{figure:Carphone2} of this
paper). Generally speaking, this algorithm can be used as an option
for further enhancing the performance of a perceptual encryption
scheme based on other techniques.

\subsection{Pazarci-Dip\c{c}in scheme}

In \cite{Turk:MPEG2Scrambling:IEEETCE2002}, Pazarci and Dip\c{c}in
proposed an MPEG-2 perceptual encryption scheme, which encrypts
the video in the RGB color space via four secret linear transforms
before the video is compressed by the MPEG-2 encoder. To encrypt
the RGB-format uncompressed video, each frame is divided into
$M\times M$ scrambling blocks (SB), which is composed of multiple
macroblocks of size $16\times 16$. Assuming the input and the
output pixel values are $x_i$ and $x_o$, respectively, the four
linear transforms are described as follows:
\begin{equation}
x_o=\begin{cases}
\alpha x_i, & (D,N)=(0,0),\\
FS-\alpha x_i, & (D,N)=(0,1),\\
FS(1-\alpha)+\alpha x_i, & (D,N)=(1,0),\\
FS-[FS(1-\alpha)+\alpha x_i], & (D,N)=(1,1),
\end{cases}\label{equation:TurkeyCipher}
\end{equation}
where $\alpha=\alpha^*/100\;(\alpha^*\in\{50,\cdots,90\})$ is a
factor controlling the visual quality degradation, $D,N$ are two
binary parameters that determine an affine transform for encryption,
and $FS$ means the maximal pixel value (for example, $FS=255$ for
8-bit RGB-videos). The value of $\alpha$ in each SB is calculated
from the preceding I-frame, with a function called $\alpha$-rule
(see Sec.~2.2 of \cite{Turk:MPEG2Scrambling:IEEETCE2002} for more
details). The $\alpha$-rule and its parameters are designated to be
the secret key of this scheme.

The main merit of the Pazarci-Dip\c{c}in scheme is that the
encryption/decryption and the MPEG encoding/decoding processes are
separated, which means that the encryption part can simply be added
to an MPEG system without any modification. However, the following
defects make this scheme problematic in real applications.

\begin{enumerate}
\item Unrecoverable quality loss caused by the encryption always
exists, unless $\alpha=1$ (which corresponds to no encryption). Even
authorized users who know the secret key cannot recover the video
with the original quality. Although it is claimed in
\cite{Turk:MPEG2Scrambling:IEEETCE2002} that human eyes are not
sensitive to such a quality loss if $\alpha$ is set above 0.5, it
may still be undesirable for high-quality video services, such as
DVD and HDTV. In addition, limiting the value of $\alpha$ lowers the
security and flexibility of the encryption scheme.

\item The compression ratio may be significantly influenced by
encryption if there are fast motions in the plain videos. This is
because the motion compensation algorithm may fail to work for
encrypted videos. The main reason is that the corresponding SBs
may be encrypted with different parameters. To reduce this kind of
influence, the encryption parameters of all SBs have to be
sufficiently close to each other. This, however, compromises the
encryption performance and the security.

\item The scheme is not suitable for encrypting MPEG-compressed
videos. In many applications, such as VoD services, the
plain-videos have already been compressed in MPEG format and
stored in digital storage media (DSM). In this case, the
Pazarci-Dip\c{c}in scheme becomes too expensive and slow, since
the videos have to be first decoded, then encrypted, and finally
encoded again. Note that the re-encoding may reduce the video
quality, since the encoder is generally different from the
original one that produced the videos in the factory. Apparently,
this defect is a natural side effect of the merit of the
Pazarci-Dip\c{c}in scheme.

\item The scheme is not secure enough against brute-force attacks.
For a given color component $C$ of any $2\times 2$ SB structure, one
can exhaustively guess the $\alpha$-values of the four SBs to
recover the $2\times 2$ SB structure, by minimizing the block
artifacts occurring between adjacent SBs. For each color component
of a SB, the value of $\alpha^*=100\alpha\in\{50,\cdots,90\}$,
$D\in\{0,1\}$ and $N$ is determined by $D$, so one can calculate
that the searching complexity is only $(41\times 2)^4\approx
2^{25.4}$, which is sufficiently small for PCs\footnote{Even when
$\alpha^*\in\{0,\cdots,100\}$, the searching complexity is only
$(101\times 2)^4\approx 2^{30.6}$, which is still practically
small.}. Once the value of $\alpha$ of an SB is obtained, one can
further break the secret key of the corresponding $\alpha$-rule. For
the exemplified $\alpha$-rule given in Eq. (3) of
\cite{Turk:MPEG2Scrambling:IEEETCE2002}, the secret key consists of
the addresses of two selected subblocks (of size $P\times P$) in a
$2\times 2$ SB structure, and a binary shift value
$\Delta\in\{5,50\}$. Because $\Delta$ can be uniquely determined
from $D$, one only needs to search other part of the key, which
corresponds to a complexity of
$\left(3\times\left(2M/P\right)^2\right)^2$. When $P=\frac{M}{2}$,
the complexity is $48^2=2304\approx 2^{11.2}$, and when when
$P=\frac{M}{4}$ it is $192^2=36864\approx 2^{15.2}$. Apparently, the
key space is not sufficiently large to resist brute-force attacks,
either. In addition, since the values of quality factors and the
secret parameters corresponding to the three color components can be
separately guessed, the whole attack complexity is only three times
of the above values, which is still too small from a cryptographical
point of view \cite{Schneier:AppliedCryptography96}. Although using
multiple secret keys for different SBs can increase the attack
complexity exponentially, the key size will be too long and the
key-management will become more complicated. Here, note that the
$\alpha$-rule itself should not be considered as part of the key,
following the well-known Kerckhoffs' principle in modern
cryptography \cite{Schneier:AppliedCryptography96}.

\item The scheme is not sufficiently sensitive to the mismatch of
the secret key, since the encryption transforms and the
$\alpha$-rule given in \cite{Turk:MPEG2Scrambling:IEEETCE2002} are
both linear functions. This means that the security against
brute-force attacks will be further compromised, as an approximate
value of $\alpha$ may be enough to recover most visual information
in the plain-video.

\item The scheme is not secure enough against
known/chosen-plaintext attacks. This is because the value of
$\alpha$ can be derived approximately from the linear relation
between the plain pixel-values and the cipher pixel-values in the
same SB. Similarly, the value $N$ can be derived from the sign of
the slope of the linear map between $x_i$ and $x_o$, and the value
of $D$ can be derived from the value range of the map.
Furthermore, assuming that there are $k$ secret parameters in the
$\alpha$-rule, if more than $k$ different values of $\alpha$ are
determined as above, it is possible to uniquely solve the
approximate values of the $k$ secret parameters. To resist
known/chosen-plaintext attacks, the secret key has to be changed
more frequently than that suggested in
\cite{Turk:MPEG2Scrambling:IEEETCE2002} (one key per program),
which will increase the computational burden of the servers
(especially the key-management system).
\end{enumerate}

\subsection{Wang-Yu-Zheng scheme}

A different scheme working in the DCT domain (between DCT transform
and Huffman entropy coding) was proposed by Wang, Yu and Zheng in
\cite{Chinese:MPEG2Scrambling:IEEETCE2003}, which can be used as an
alternative solution to overcome the first two shortcomings of the
Pazarci-Dip\c{c}in scheme. By dividing all 64 DCT coefficients of
each $8\times 8$ block into 16 sub-bands following the distance
between each DCT coefficient and the DC coefficient, this new scheme
encrypts the $j$-th AC coefficient in the $i$-th sub-band as
follows:
\begin{equation}
b_{ij}'=\begin{cases}
b_{ij}-\lfloor\beta a_i\rfloor, & b_{ij}\geq 0,\\
b_{ij}+\lfloor\beta a_i\rfloor, & b_{ij}<0,
\end{cases}
\end{equation}
where $b_{ij}$ and $b_{ij}'$ denotes the plain pixel-value and the
cipher pixel-value, respectively, $\beta\in[0,1]$ is the control
factor, $a_i$ is the rounding average value of all AC coefficients
in the $i$-th sub-band, and $\lfloor\cdot\rfloor$ means the rounding
function towards zero. The DC coefficients are encrypted in a
different way, as $b_0'=b_0\pm\lfloor C\beta a_0\rfloor$, where
$a_0=b_0$ and $C\in[0,1]$ is the second control factor\footnote{Note
that the rounding function is missed in Eqs. (3) and (4) of
\cite{Chinese:MPEG2Scrambling:IEEETCE2003}. In addition, Eq. (4) of
\cite{Chinese:MPEG2Scrambling:IEEETCE2003} should read
$b_0'=b_0\pm\lfloor C\beta a_0\rfloor$, not $b_0'=a_0\pm C\beta$.}.
The value of $a_i$ can also be calculated in a more complicated way
to enhance the encryption performance, following Eqs. (5) and (6) in
\cite{Chinese:MPEG2Scrambling:IEEETCE2003}, where three new
parameters, $k_1,k_2,k_3$ are introduced to determine the values of
$a_i$ for the three color components, Y, Cr and Cb. The 16 average
values, $a_0\sim a_{15}$, the two control factors, $\beta$ and $C$,
and the three extra parameters (if used), $k_1,k_2,k_3$, altogether
serve as the secret scrambling parameters (i.e., the secret key) of
each SB. Three different ways are suggested for the transmission of
the secret parameters: a) encrypting them and transmitting them in
the payload of TS (transport stream); b) embedding them in the
high-frequency DCT coefficients; c) calculating them from the
previous I-frame in a way similar to the $\alpha$-rule in
\cite{Turk:MPEG2Scrambling:IEEETCE2002}.

In fact, the Wang-Yu-Zheng scheme is just an enhanced version of the
Pazarci-Dip\c{c}in scheme, without amending all shortcomings of the
latter scheme. Precisely, the following problems still remain.
\begin{enumerate}
\item Though the reduction of the compression ratio about motion
compensations is avoided, the encryption will change the natural
distribution of the DCT coefficients and thus reduce the
compression efficiency of the Huffman entropy encoder. For
example, when each sub-band has only one non-zero coefficient, it
is possible that all 64 coefficients become non-zero after the
encryption. This significantly increases the video size. In
addition, if the secret parameters are embedded into the
high-frequency DCT coefficients for transmission, the compression
performance will be further compromised.

\item The scheme is still not sufficiently sensitive to the
mismatch of the secret parameters, since the encryption function and
the calculation function of $a_i$ are kept linear. It is still not
sufficiently secure against brute-force attacks to the secret
parameters, because of the limited values of
$a_i,\beta,C,k_1,k_2,k_3$. Furthermore, due to the non-uniform
distribution of the DCT coefficients in each sub-band, an attacker
needs not to randomly search all possible values of $a_i$.

\item This scheme is still insecure against known/chosen-plaintext
attacks if the third way is used for calculating the secret
parameters. In this case, $a_i$ of each SB can be easily calculated
from the previous I-frame of the plain-video. Additionally, since
the value of $\lfloor\beta a_i\rfloor$ can be obtained from
$b_{ij}-b_{ij}'$, the secret parameter $\beta$ can be derived
approximately. In a similar way, the secret parameter $C$ can also
be derived approximately. If $a_i$ is calculated with $k_1,k_2,k_3$,
the values of $\beta,k_1,k_2,k_3$ can be solved approximately with a
number of known/chosen AC coefficients in four or more different
sub-bands, so that $C$ can be further derived from one known/chosen
DC coefficient.

\item The method of transmitting the secret parameters in the
payload of the transport stream cannot be used under the following
conditions: a) the key-management system is not available; b) the
video is not transmitted with the TS format. A typical example is
the perceptual encryption of MPEG-video files in personal
computers.

\end{enumerate}

\section{More Efficient Design of Perceptual MPEG-Video Encryption Schemes}
\label{section:NewDesign}

Based on the analysis given above, we propose a simpler design of
perceptual encryption for MPEG videos, and attempt to overcome the
problems in existing schemes. The following useful features are
supported in our new design.
\begin{itemize}
\item \textit{Format-compliance}: the encrypted video can still be
decoded by any standard-compliant MPEG decoder. This is a basic
feature of all perceptual encryption schemes.

\item \textit{Lossless visual quality}: the encrypted video has
the same visual quality as the original one, i.e., the original
full-quality video can be exactly recovered when the secret key is
presented correctly.

\item \textit{Strict size-preservation}: the size of each data
element in the lowest syntax level, such as VLCs, FLCs and
continuous stuffing bits, remains unchanged after encryption. When
the video stream is packetized in a system stream (i.e., PS or
TS), the size of each video packet remains unchanged after
encryption. This enables the following useful features in
applications:
\begin{itemize}
\item \textit{independence of bit-rate types (VBR and FBR)};

\item \textit{avoiding some time-consuming operations when
encrypting MPEG-compressed videos}: bit-rate control,
re-packetization of the system stream and the re-multiplexing of
multiple audio/video streams;

\item \textit{on-the-fly encryption}: a) direct encryption of
MPEG-compressed video files without creating temporary files, i.e.,
one can open an MPEG video file, read the bitstream and
simultaneously update (encrypt) it; b) instantaneous switching
encryption on/off for online video transmission;

\item \textit{ROI (region on interest) encryption}: selectively
encrypting partial frames, slices, macroblocks, blocks, motion
vectors and/or DCT coefficients within specific regions of the
video.
\end{itemize}

\item \textit{Independence of optional data elements in MPEG
videos}: the perceptibility is not obtained by encrypting optional
data elements, such as quantiser\_matrix and
coded\_block\_pattern\footnote{Strictly speaking, motion vectors
are also optional elements in MPEG videos, so the scheme should
not encrypt only motion vectors as did in
\cite{Yann:WaterScrambling:ACMMM2002}.}. This means that the
scheme can encrypt any MPEG-compliant videos with a uniform
performance.

\item \textit{Fast encryption speed}: a) the extra computational
load added by the encryption is much smaller than the
computational load of a typical MPEG encoder; b) MPEG-compressed
videos can be quickly encrypted without being fully decoded and
re-encoded (at least the time-consuming IDCT/DCT operations are
avoided).

\item \textit{Easy implementation}: the encryption/decryption
parts can be easily incorporated into the whole MPEG system,
without major modification of the structure of the codec.

\item \textit{Multi-dimensional perceptibility}: the degradation
of visual quality is controlled by multi-dimensional factors.

\item \textit{Security against known/chosen-plaintext attacks} is
ensured by four different measures.
\end{itemize}
To the best of our knowledge, some of the above features (such as
on-the-fly encryption) have never been discussed in the literature
on video encryption, in spite of their usefulness in real
applications.

With the above features, the perceptual encryption scheme becomes
more flexible to fulfill different requirements of various
applications. To realize the strict size-preservation feature, the
encryption algorithm has to be incorporated into the MPEG encoder,
i.e., the (even partial) separation of the cipher and the encoder is
impossible. This is a minor disadvantage in some applications.
However, if the re-design is sufficiently simple, it is worth doing
so to get a better tradeoff between the overall performance and the
easy implementation. In the case that the re-design of the MPEG
codec is impossible, for example, if the codec is secured by the
vendor, a simplified MPEG codec can be developed for the embedding
of the perceptual video cipher. Since the most time-consuming
operations in a normal MPEG codec including DCT/IDCT and picture
reconstruction, are excluded from the simplified MPEG codec, fast
encryption speed and low implementation complexity of the whole
system can still be achieved.

In the following text of this section, we describe the design
principle along with different methods of providing security
against known/chosen-plaintext attacks, and discuss several
implementation issues.

\subsection{The new design}

This design is a generalized version of VEA
\cite{Shi:MPEGEncryption:MMTA2004} for perceptual encryption, by
selectively encrypting FLC data elements in the video stream.
Apparently, encrypting FLC data elements is the most natural and
perhaps the simplest way to maintain all needed features,
especially the need for the strict size-preservation feature. The
proposed scheme is named PVEA -- perceptual video encryption
algorithm. Note that PVEA can also be considered as an enhanced
combination of the encryption techniques for JPEG images proposed
in \cite{Belgian:SelectiveImageEncryption:ACIVS2002,
Torrubia:PerceptualJPEG:ICCE2003} and the perceptual encryption of
motion vectors \cite{Yann:WaterScrambling:ACMMM2002}.

There are three main reasons for selecting only FLC data elements
for encryption.
\begin{enumerate}
\item
As analyzed below, all existing VLC encryption algorithms cannot be
directly used to provide a controllable degradation of the quality.
New ideas have to be developed to adopt VLC encryption in perceptual
encryption schemes.
\begin{itemize}
\item
\textit{VLC encryption with different Huffman tables}
\cite{Wu&Kuo:AudiovisualEncryption:SPIE2001,
Wu&Kuo:EntropyCodecEncryption:SPIE2001,
Xie&Kuo:MHTEncryption:SPIE2003, Wu&Kuo:MHTEncryption:IEEETMM2004,
Kankanhalli:VideoScrambler:IEEETCE2002,
Shi:SecretHuffmanCoding:MM98, Shi:MPEGEncryption:MMTA2004}: Since
each VLC-codeword is a pair of (run, level), if a VLC-codeword is
decoded to get an incorrect ``run" value, then the position of all
the following DCT coefficients will be wrong. As a result, the
visual quality of the decoded block will be degraded in an
uncontrollable way. Thus, it is difficult to find a factor to
control such visual quality degradation. Moreover, if the Huffman
tables do not keep the size of each VLC-entry as designated in
\cite{Wu&Kuo:AudiovisualEncryption:SPIE2001,
Wu&Kuo:EntropyCodecEncryption:SPIE2001,
Xie&Kuo:MHTEncryption:SPIE2003, Wu&Kuo:MHTEncryption:IEEETMM2004,
Kankanhalli:VideoScrambler:IEEETCE2002}, syntax errors may occur
when an unauthorized user decodes an encrypted video. This means
that the encryption cannot ensure the format compliance to any
standard MPEG codecs.

\item
\textit{VLC-index encryption} \cite{Zeng:VideoScrambling:MMSP2001,
Zeng:VideoScrambling:IEEETCASVT2002}: This encryption scheme can
ensure format compliance, but still suffers from the
uncontrollability of the visual quality degradation due to the same
reason as above. Another weakness of VLC-index encryption is that it
may influence the compression efficiency and bring overhead on video
size.

\item
\textit{Shuffling VLC-codewords or RLE events before the entropy
encoding stage} \cite{Zeng:VideoScrambling:IEEETCASVT2002,
Liu:MPEGEncryption:IEICE-A2006}: This algorithm can ensure both the
format compliance and the strict size-preservation. However, even
exchanging only two VLC-codewords may cause a dramatic change of the
DCT coefficients distribution of each block. So, this encryption
algorithm cannot realize a slight degradation of the visual quality
and fails to serve as an ideal candidate for perceptual encryption.
\end{itemize}

\item
It is obvious that FLC encryption is the simplest way to achieve all
the desired properties mentioned in the beginning of this section,
especially to achieve format compliance, strict size-preservation
and fast encryption \textbf{simultaneously}. For example, naive
encryption\footnote{In the image/video encryption literature, the
term ``naive encryption" means to consider the video as a 1-D
bitstream and encrypt it via a common cipher.} can realize strict
size-preservation and fast encryption, but cannot ensure format
compliance.

\item
As will be seen below, using FLC encryption is sufficient to fulfill
the needs of most real applications for perceptual encryption.
\end{enumerate}

According to MPEG standards \cite{MPEG1-ISOStandard,
MPEG2-ISOStandard, MPEG4-ISOStandard}, the following FLC data
elements exist in an MPEG-video bitstream:
\begin{itemize}
\item 4-byte start codes: 000001xx (hexadecimal);

\item almost all information elements in various headers;

\item sign bits of non-zero DCT coefficients;

\item (differential) DC coefficients in intra blocks;

\item ESCAPE DCT coefficients;

\item sign bits and residuals of motion vectors.
\end{itemize}

To maintain the format-compliance to the MPEG standards after the
encryption, the first two kinds of data elements should not be
encrypted. So, in PVEA, only the last four FLC data elements are
considered, which are divided into three categories according to
their contributions to the visual quality:
\begin{itemize}
\item \textit{intra DC coefficients}: corresponding to the rough
view (in the level of $8\times 8$ block) of the video;

\item \textit{sign bits of non-intra DC coefficients and AC
coefficients, and ESCAPE DCT coefficients}: corresponding to
details in $8\times 8$ blocks of the video;

\item \textit{sign bits and residuals of motion vectors}:
corresponding to the visual quality of the video related to the
motions (residuals further corresponds to the details of the
motions).
\end{itemize}
Based on the above division, three control factors, $p_{sr}$,
$p_{sd}$, and $p_{mv}$ in the range [0,1], are used to control the
visual quality in three different dimensions: the low-resolution
rough (spatial) view, the high-resolution (spatial) details, and
the (temporal) motions. With the three control factors, the
encryption procedure of PVEA can be described as follows:
\begin{enumerate}
\item encrypting intra DC coefficients with probability $p_{sr}$;

\item encrypting sign bits of non-zero DCT coefficients (except
for intra DC coefficients) and ESCAPE DCT coefficients with
probability $p_{sd}$;

\item encrypting sign bits and residuals of motion vectors with
probability $p_{mv}$.
\end{enumerate}
The encryption of selected FLC data elements can be carried out with
either a stream cipher or a block cipher. When a block cipher is
adopted, the consecutive FLC data elements should be first
concatenated together to form a longer bit stream, then each block
of the bit stream is encrypted, and finally each encrypted FLC data
element is placed back into its original position in the video
stream. Under the assumption that the stream cipher or block cipher
embedded in PVEA is secure, some special considerations should be
taken into account in order to ensure the security against various
attacks, as discussed below.

In the above-described PVEA, the three factors control the visual
quality, as follows:
\begin{itemize}
\item $p_{sr}=1\to 0$: the spatial perceptibility changes from
``almost imperceptible" to ``perfectly perceptible" when
$p_{sd}=0$ or to ``roughly perceptible" when $p_{sd}>0$;

\item $p_{sr}=0$, $p_{sd}=1\to 0$: the spatial perceptibility
changes from ``roughly perceptible" to ``perfectly perceptible";

\item $p_{mv}=1\to 0$: the temporal (motion) perceptibility (for
P/B-pictures only) changes from ``almost imperceptible" to
``perfectly perceptible".
\end{itemize}
The encryption may bring the recovered motion vectors out of the
spatial range of the picture, so the motion compensation
operations (or even the involved picture itself) may be simply
discarded by the MPEG decoder. In this case, the temporal (motion)
perceptibility will be ``perfectly imperceptible", not just
``almost imperceptible".

In the Appendix of \cite{Shi:MPEGEncryption:MMTA2004}, it was
claimed that the DC coefficients of each block can be uniquely
derived from the other 63 AC coefficients. This means that the
perceptual encryption of DC coefficients must not be used alone,
i.e., some AC coefficients must also be encrypted to make the
encryption of the DC coefficients secure. It was lately observed
that this claim is not correct
\cite{Li:MPEGEncryption:MMTA2004note}. In fact, the DC coefficient
of a block means the average brightness of the block, and is
independent of the other 63 AC coefficients. Thus, the
DC-encryption and AC-encryption of PVEA are independent of each
other, i.e., the two control factors, $p_{sr}$ and $p_{sd}$, are
independent of each other, and they can be freely combined in
practice.

\subsection{Security against ciphertext-only attacks and a constraint of the control
factor}

The format compliance of perceptual encryption makes it possible for
the attacker to guess the values of all encrypted FLC data elements
separately in ciphertext-only attacks. The simplest attack is to try
to recover more visual information by setting all the encrypted FLC
data elements to zeros. This is called error-concealment-based
attack (ECA) \cite{Zeng:VideoScrambling:IEEETCASVT2002}. Our
experimental results have shown that PVEA is secure against such
attack. More details are given in the next section.

To guess the value of each FLC data element, one can also employ the
local correlation existing between adjacent blocks in each frame.
That is, one can search for a set of all encrypted FLC data elements
in each frame to achieve the least blocking artifact. Does such a
deblocking attack work? Now let us try to get a lower bound of this
attack's complexity, by assuming that the number of all FLC data
elements in each frame is $N$, which means that the number of
encrypted FLC data elements is $pN$. Then, the complexity of the
deblocking attack will not be less than
$O\left(\binom{N}{pN}2^{pN}\right)$, since each FLC data elements
has at least two candidate values. So, if $\binom{N}{pN}2^{pN}$ is
cryptographically large, the deblocking attack will not compromise
the security of PVEA. As a lower bound of $p$ corresponding to a
typical security level, one can get $p\geq 100/N$ by assuming
$2^{pN}\geq 2^{100}$. For most consumer videos that need to be
protected via perceptual encryption, $N$ is generally much larger
than 100, so this constraint on $p$ generally does not have too much
influence on the overall performance of PVEA. Because the complexity
$2^{pN}$ is much over-estimated\footnote{There are two reasons about
the over-estimation: 1) the omission of $\binom{N}{pN}$, which is
very large when $N\gg pN$ and $pN$ is not very small; 2) some FLC
elements (such as intra DC coefficients) have more than 2 candidate
values.}, the constraint can be further relaxed in practice. For
example, when $N=200$, the above condition suggests that $p\geq
100/N=1/2$. However, calculations showed that $p\geq 9/100$ is
enough to ensure a complexity larger than $O(2^{100})$.

Since it is generally impractical to carry out the deblocking attack
on the whole frame, another two-layer deblocking attack may be
adopted by the attacker: 1) performing the deblocking attack on
small areas of the frame; 2) for all candidates of these small
areas, performing the deblocking attack on the area-level again.
Though this two-layer attack generally has a much smaller complexity
than the simple attack, its efficiency is still limited due to the
following reasons.
\begin{itemize}
\item
For each small area, the number of encrypted FLC elements is
generally not equal to $pN^*$, where $N^*$ denotes the total number
of all FLC elements in the area. Thus, even this number has to be
exhaustively guessed and then validated by considering the numbers
of other areas (i.e., the whole frame). The existence of three
independent quality factors makes the attack even more complicated.

\item
For small each areas, the probability that the least deblocking
result does not correspond to the real scene may not very small.
Accordingly, the attacker has to mount a more loose deblocking
attack, thus leading to a higher attacking complexity.

\item
Even for the smallest area of size $16\times 16$, there are
generally more than one hundred FLC elements (i.e., $N^*\geq 100$),
especially when there are rich visual information included in the
area.

\item
If the number of FLC elements in an area is relatively small, this
area generally contains less significant visual information (such as
a smooth area).

\item
The smaller each area is, the more the number of fake results will
be, and then the more the complexity of the second stage will be.
\end{itemize}
Of course, with the two-layer deblocking attack, the attacker can
have a chance to recover a number of small areas, though he/she
generally cannot get the whole frame. Such a minor security problem
is an unavoidable result of the inherent format-compliance property
of the perceptual encryption algorithms and related to the essential
disadvantage of perceptual encryption exerted on some special
MPEG-videos (see the discussion on Fig.~\ref{figure:Animation} in
the next section).

\subsection{Security against known/chosen-plaintext attacks}

Generally speaking, there are four different ways to provide
security against known/chosen-plaintext attacks. Users can select
one solution for a specific application.

\subsubsection{Using a block cipher}

With a block cipher, it is easy to provide security against
known/chosen-plaintext attacks. Since the lengths of different FLC
data elements are different, the block cipher may have to run in CFB
(cipher feedback) mode with variable-length feedback bits to realize
the encryption. Note that $n$-bit error propagation exists in block
ciphers running in the CFB mode
\cite{Schneier:AppliedCryptography96}, where $n$ is the block size
of the cipher. It is also possible to cascade multiple FLC data
elements to compose an $n$-bit block for encryption, as in RVEA
\cite[Sec. 7]{Shi:MPEGEncryption:MMTA2004}. Compared to the CFB
mode, the latter encryption mode can achieve a faster encryption
speed (with a little more implementation complexity for bit
cascading), since in the CFB mode only one element can be encrypted
in each run of the block cipher.

\subsubsection{Using a stream cipher with plaintext/ciphertext
feedback}

After encrypting each plain data element, the plaintext or the
ciphertext is sent to perturb the stream cipher for the encryption
of the next plain data element. In such a way, the keystream
generated by the stream cipher becomes dependent on the whole
plain-video, which makes the known/chosen-plaintext attacks
impractical. Note that an initial vector is needed for the
encryption of the first plain data element.

\subsubsection{Using a key-management system and a stream cipher}

When a key-management system is available in an application, the
encryption procedure of PVEA can be realized with a stream cipher.
To effectively resist known/chosen-plaintext attacks, the secret key
of the stream cipher should be frequently changed by the
key-management system. In most cases, it is enough to change one key
per picture, or per GOP. Note that this measure needs more
computational load with higher implementation cost, and is suitable
mainly for encrypting online videos.

\subsubsection{Using a stream cipher with UID}

When key-management systems are not available in some applications,
a unique ID (UID) can be used to provide the security against
known/chosen-plaintext attacks by ensuring that the UIDs are
different for different videos. The UID of an MPEG-video can be
stored in the user\_data area. The simplest form of the UID is the
vendor ID plus the time stamp of the video. It is also possible to
determine the UID of a video with a hash function or a secure
pseudo-random number generator (PRNG). In this case, the UIDs of two
different videos may be identical, but the probability is
cryptographically small if the UID is sufficiently long. The UID is
used to initialize the stream cipher together with the secret key,
which ensures that different videos are encrypted with different
keystreams. Thus, when an attacker successfully gets the keystream
used for $n$ known/chosen videos, he cannot use the broken
keystreams to break other different videos. Of course, the employed
stream cipher should be secure against plaintext attacks in the
sense that the secret key cannot be derived from a known/chosen
segment of the long keystream that encrypts the whole video stream
\cite{Schneier:AppliedCryptography96}.

\subsection{Implementation issues}

Since PVEA is a generalization of VEA, it is obvious that fast
encryption speed can be easily achieved, as shown in
\cite{Shi:MPEGEncryption:MMTA2004}. In addition, by carefully
optimizing the implementation, the encryption speed can be further
increased. We give two examples to show how to optimize the
implementation of PVEA so as to increase the encryption speed.

A typical way to realize the probabilistic quality control with a
decimal factor $p$ is as follows: generate a pseudo-random
decimal, $r\in[0,1]$, for each data element with a
uniformly-distributed PRNG, and then encrypt the current element
only when $r\leq p$. The above implementation can be modified as
follows to further increase the encryption speed:
\begin{enumerate}
\item pseudo-randomly select $N_p=\mathrm{round}(N\cdot p)$
integers from the set $\{0,\cdots,N-1\}$;

\item create a binary array $SE[0]\sim SE[N]$: $SE[i]=1$ if the
integer $i$ is selected; otherwise, $SE[i]=0$;

\item encrypt the $i$-th FLC data element only when $SE[i\bmod
N]=1$.
\end{enumerate}
In this modified implementation, only a modulus addition and a
look-up-table operation are needed to determine whether the current
data element should be encrypted. As a comparison, in the typical
implementation, one run of the PRNG is needed for each data element,
which is generally much slower. Although $N$ bits of extra memory is
needed to store the array in the modified implementation, it is
merely a trivial problem since video codec generally requires much
more memory. To ensure the security against deblocking attacks, in
the modified implementation the value of $N$ should not be too
small\footnote{In most cases, it is enough to set $N\geq 300$.}.

To further reduce the computational load of PVEA, another way is to
selectively encrypt partial FLC data elements. Two possible options
are as follows: 1) encrypt only intra blocks; 2) encrypt only sign
bits (or a few number of most significant bits) of intra DC
coefficients, ESCAPE DCT coefficients, and residuals of motion
vectors. The above two options can also be combined together. This
will have very little effect on the encryption performance, since an
attacker can only recover video frames with a poor visual quality
from other unencrypted data elements
\cite{Agi&Gong:StudySecureMPEG:NDSS96,
Zeng:VideoScrambling:IEEETCASVT2002}.

\section{Encryption Performance of PVEA}
\label{section:Experiments}

\begin{figure}
\centering \centering
\begin{minipage}{\figwidth}
\centering
\includegraphics[width=\textwidth]{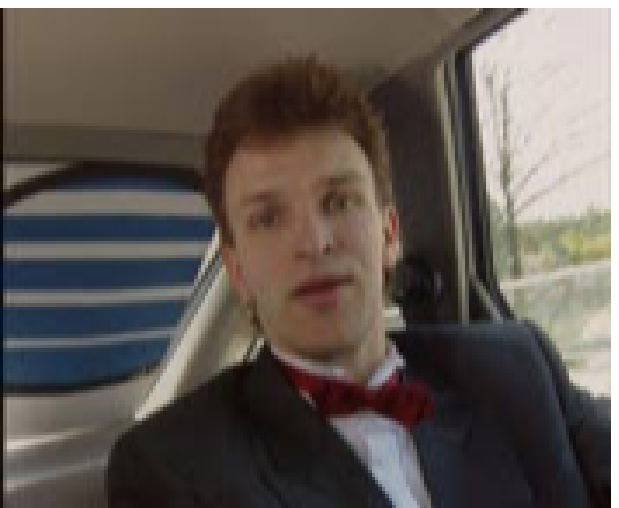}
a)
\end{minipage}
\begin{minipage}{\figwidth}
\centering
\includegraphics[width=\textwidth]{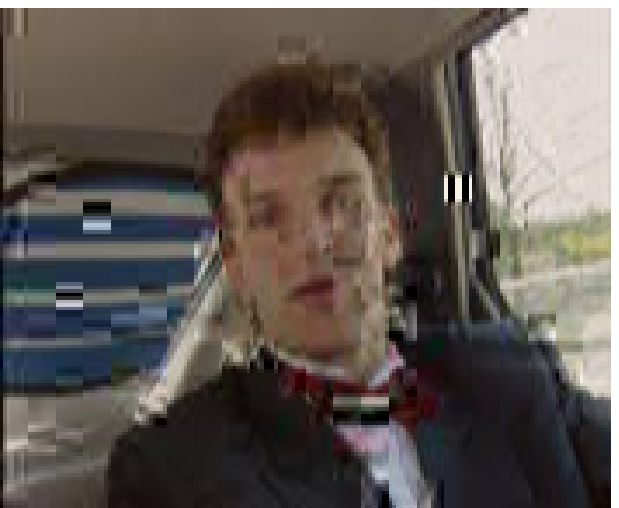}
 b)
\end{minipage}
\begin{minipage}{\figwidth}
\centering
\includegraphics[width=\textwidth]{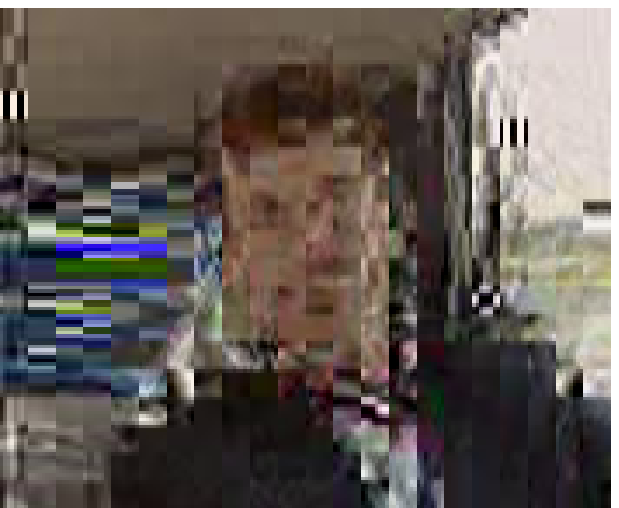}
c)
\end{minipage}
\begin{minipage}{\figwidth}
\centering
\includegraphics[width=\textwidth]{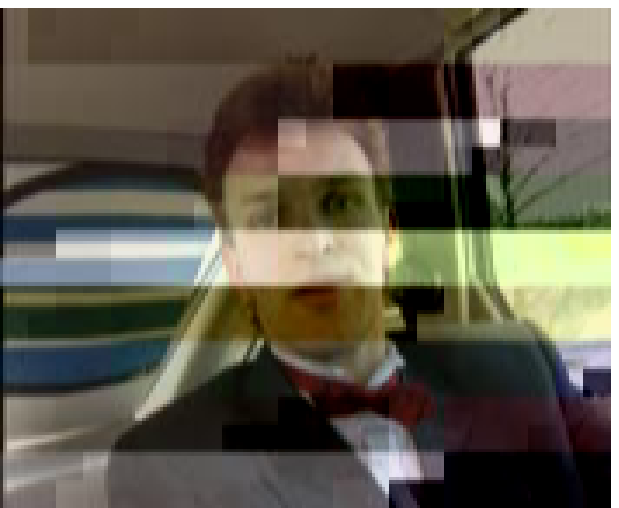}
d)
\end{minipage}
\begin{minipage}{\figwidth}
\centering
\includegraphics[width=\textwidth]{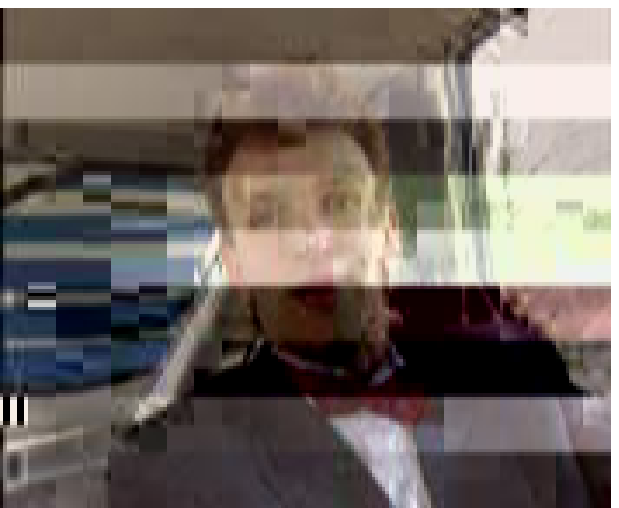}
e)
\end{minipage}
\begin{minipage}{\figwidth}
\centering
\includegraphics[width=\textwidth]{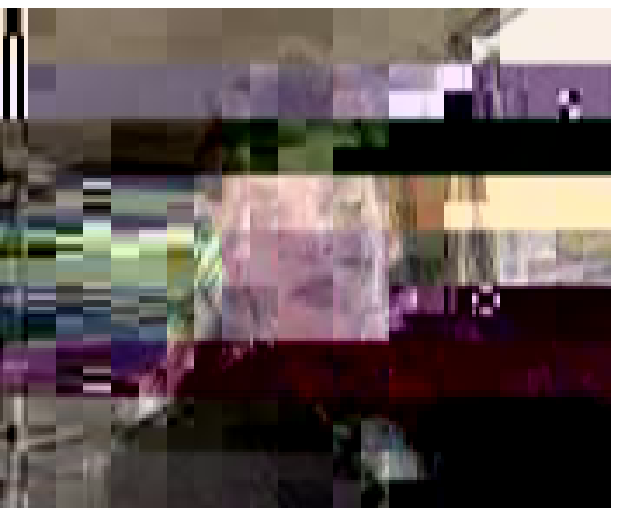}
f)
\end{minipage}
\begin{minipage}{\figwidth}
\centering
\includegraphics[width=\textwidth]{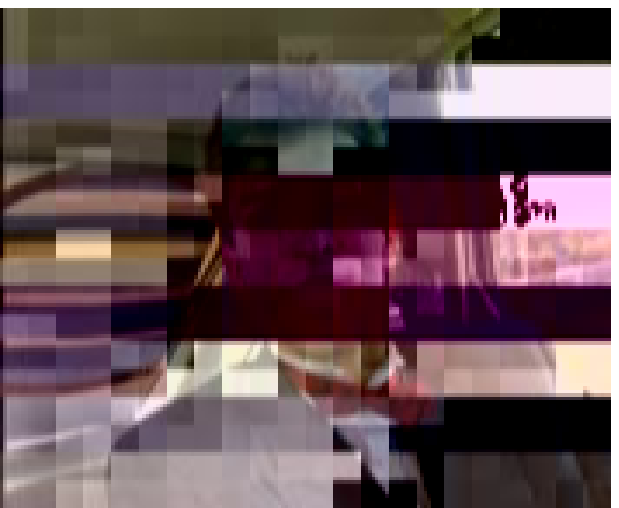}
g)
\end{minipage}
\begin{minipage}{\figwidth}
\centering
\includegraphics[width=\textwidth]{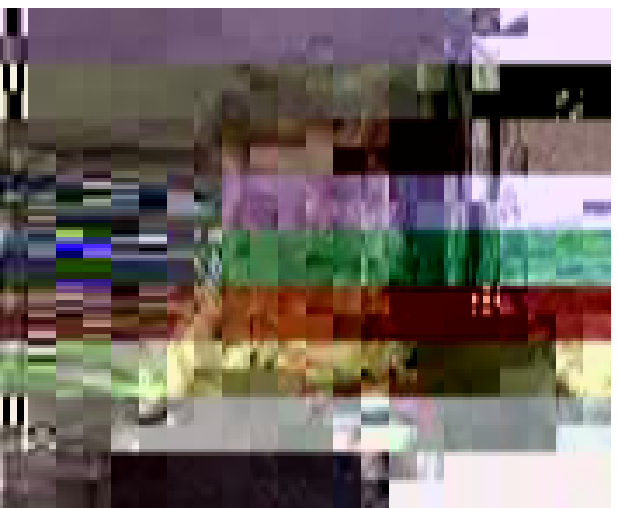}
h)
\end{minipage}
\begin{minipage}{\figwidth}
\centering
\includegraphics[width=\textwidth]{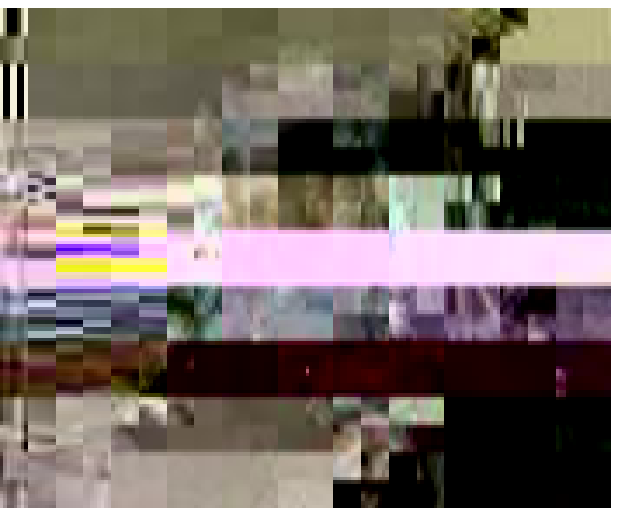}
i)
\end{minipage}
\caption{The encryption results of the 1st frame in ``Carphone":
a) $(p_{sr},p_{sd})=(0,0)$ -- the plain frame; b)
$(p_{sr},p_{sd})=(0,0.2)$; c) $(p_{sr},p_{sd})=(0,1)$; d)
$(p_{sr},p_{sd})=(0.2,0)$; e) $(p_{sr},p_{sd})=(0.2,0.2)$; f)
$(p_{sr},p_{sd})=(0.5,0.5)$; g) $(p_{sr},p_{sd})=(1,0)$; h)
$(p_{sr},p_{sd})=(1,0.2)$; i)
$(p_{sr},p_{sd})=(1,1)$.}\label{figure:Carphone}
\end{figure}

\begin{figure}
\centering
\begin{minipage}{\figwidth}
\centering
\includegraphics[width=\textwidth]{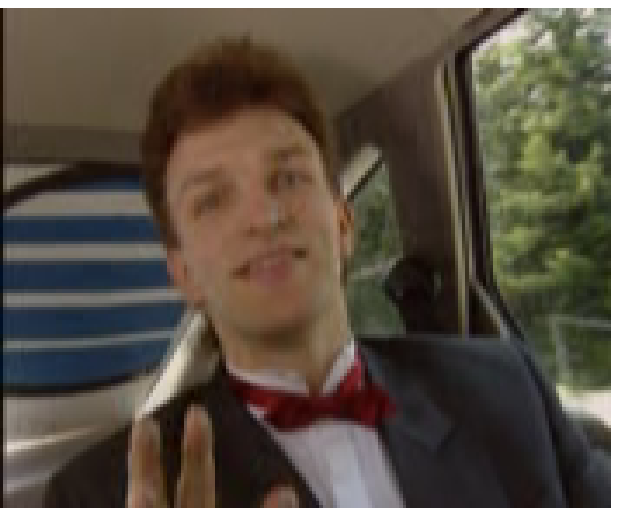}
a)
\end{minipage}
\begin{minipage}{\figwidth}
\centering
\includegraphics[width=\textwidth]{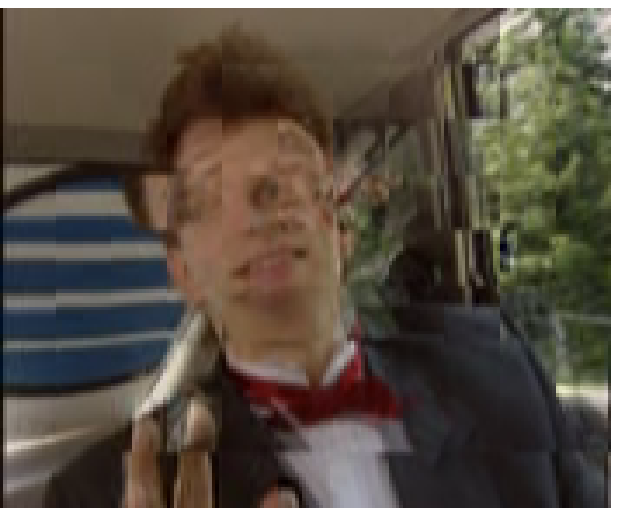}
b)
\end{minipage}
\begin{minipage}{\figwidth}
\centering
\includegraphics[width=\textwidth]{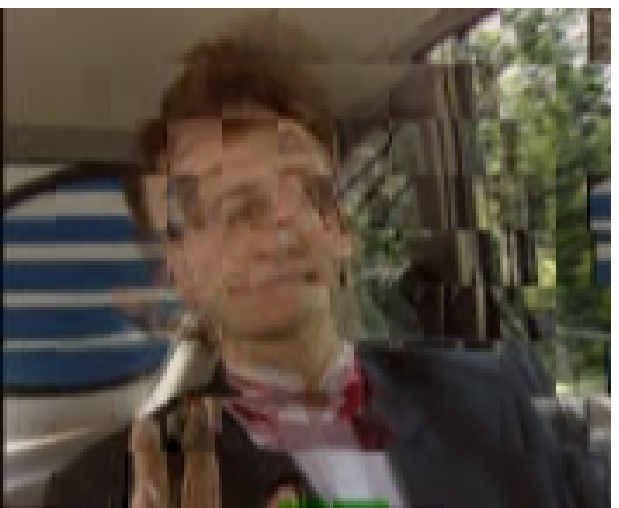}
c)
\end{minipage}
\caption{The encryption results of the 313th frame in ``Carphone":
a) $(p_{sr},p_{sd},p_{mv})=(0,0,0)$ -- the plain frame; b)
$(p_{sr},p_{sd},p_{mv})=(0,0,0.5)$; c)
$(p_{sr},p_{sd},p_{mv})=(0,0,1)$.}\label{figure:Carphone2}
\end{figure}

Some experiments have been conducted to test the real encryption
performance of PVEA for a widely-used MPEG-1 test video,
``Carphone". The encryption results of the 1st frame (I-type) are
shown in Fig.~\ref{figure:Carphone}, with different values of the
two control factors $p_{sr}$ and $p_{sd}$. It can be seen that the
degradation of the visual quality is effectively controlled by the
two factors. The encryption results of the third control factor
$p_{mv}$ are given in Fig.~\ref{figure:Carphone2}, where the 313th
frame (B-type) is selected for demonstration. It can be seen that
encrypting only the motion vectors will not cause much degradation
in the visual quality.

\begin{figure}
\centering
\begin{minipage}{\figwidth}
\centering
\includegraphics[width=\textwidth]{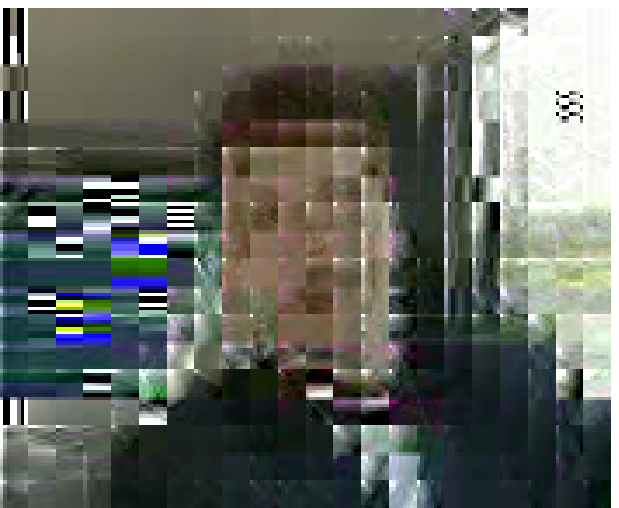}
a)
\end{minipage}
\begin{minipage}{\figwidth}
\centering
\includegraphics[width=\textwidth]{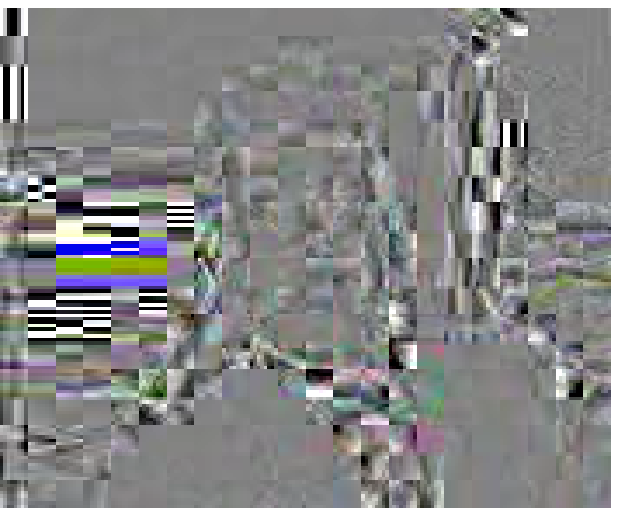}
b)
\end{minipage}
\caption{The recovered results after applying ECA for the 1st
frame in ``Carphone": a) breaking Fig.~\ref{figure:Carphone}c; b)
breaking Fig.~\ref{figure:Carphone}i.}\label{figure:Carphone3}
\end{figure}

Our experiments have also shown that PVEA is secure against
error-concealment based attacks. For two encrypted frames shown in
Fig.~\ref{figure:Carphone}, the recovered images after applying ECA
are shown in Fig.~\ref{figure:Carphone3}. In
Fig.~\ref{figure:Carphone3}a, the sign bits of all AC coefficients
are set to be zeros, and in Fig.~\ref{figure:Carphone3}b all DC
coefficients are also set to be zeros. It can be seen that the
visual quality of the recovered images via such an attack is even
worse than the quality of the cipher-images, which means that ECA
cannot help an attacker get more visual information. Actually, the
security of PVEA against ECA depends on the fact that an attacker
cannot tell encrypted data elements from un-encrypted ones without
breaking the key. As a result, he has to set all possible data
elements to be fixed values, which is equivalent to perceptual
encryption with the control factor 1, i.e., the strongest level of
perceptual encryption.

Finally, it is worth mentioning that PVEA has a minor disadvantage
that the degradation in the visual quality is dependent on the
amplitudes of the intra DC coefficients. As an extreme example,
consider an intra picture whose DC coefficients are all zeros, which
means that the FLC-encoded differential value of each intra DC
coefficient does not occur in the bitstream, i.e., only the
VLC-encoded dct\_dc\_size=0 occurs. In this case, the control of the
rough visual quality by $p_{sr}$ completely disappears. Similarly,
when dct\_dc\_size=1, the encryption can only change the
differential value from $\pm 1$ to $\mp 1$, so the degradation will
not be very significant. As a result, this problem will cause the
perceptibility of some encrypted videos become ``partially
perceptible" when $p_{sr}=1$ (should be ``almost imperceptible" for
most videos). For an MPEG-1 video\footnote{Source of this test
video:
\url{http://www5.in.tum.de/forschung/visualisierung/duenne_gitter/DG_4.mpg}.}
with a dark background (i.e., with many intra DC coefficients of
small amplitudes), the encryption results are shown in
Fig.~\ref{figure:Animation}. Fortunately, this problem is not so
serious in practice, for the following reasons:
\begin{itemize}
\item most consumer videos contain
sufficiently many intra DC coefficients of large amplitudes;

\item even when there are many zero intra DC coefficients, the
content of the video has to be represented by other intra DC
coefficients of sufficiently large amplitudes;

\item the differential encoding can increase the number of
non-zero intra DC coefficients;

\item the partial degradation caused by $p_{sr}$ and the
degradation caused by $p_{sd}$ and $p_{mv}$ are enough for most
applications of perceptual encryption (see Figs.
\ref{figure:Carphone} and \ref{figure:Animation}).
\end{itemize}

From this minor disadvantage of PVEA, a natural result can be
immediately derived: for the protection of MPEG videos that are
highly confidential, VLC data elements should also be encrypted. In
fact, our additional experiments on various video encryption
algorithms have shown that it might be impossible to effectively
degrade the visual quality of the MPEG videos with dark background
via format-compliant encryption, unless the compression ratio and
the strict size-preservation feature are compromised. The relations
among the encryption performance, the compression ratio, the
size-preservation feature, and other features of video encryption
algorithms, are actually much more complicated. These problems will
be investigated in our future research.

\setlength\figwidth{0.32\columnwidth}
\begin{figure}
\centering
\begin{minipage}{\figwidth}
\centering
\includegraphics[width=\textwidth]{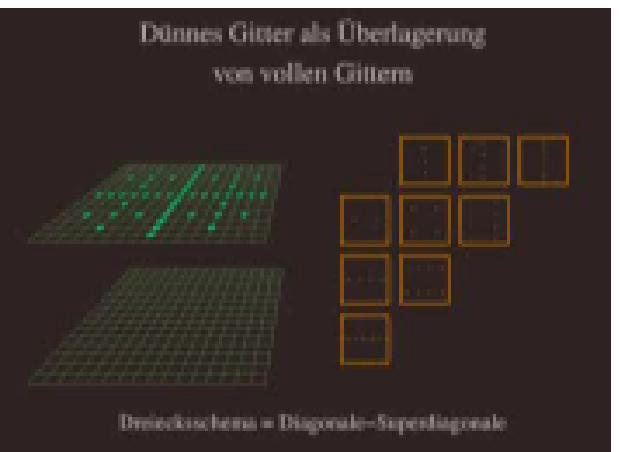}
a)
\end{minipage}
\begin{minipage}{\figwidth}
\centering
\includegraphics[width=\textwidth]{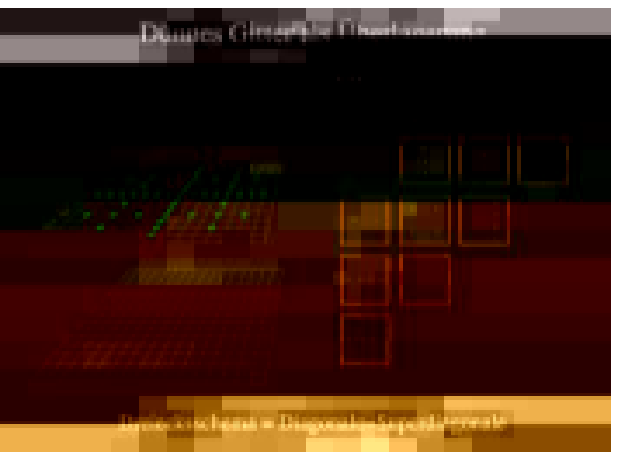}
b)
\end{minipage}\\
\begin{minipage}{\figwidth}
\centering
\includegraphics[width=\textwidth]{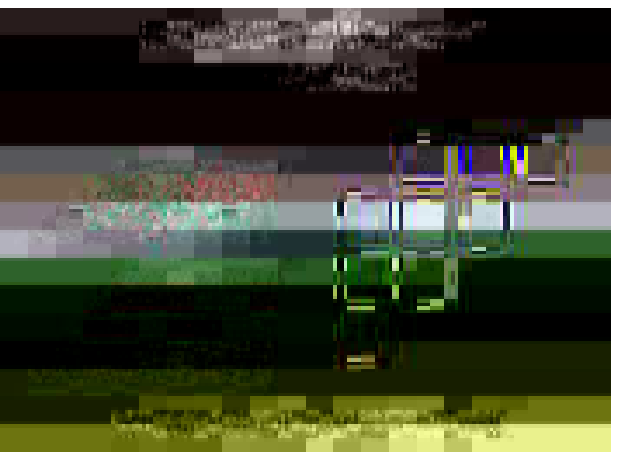}
c)
\end{minipage}
\begin{minipage}{\figwidth}
\centering
\includegraphics[width=\textwidth]{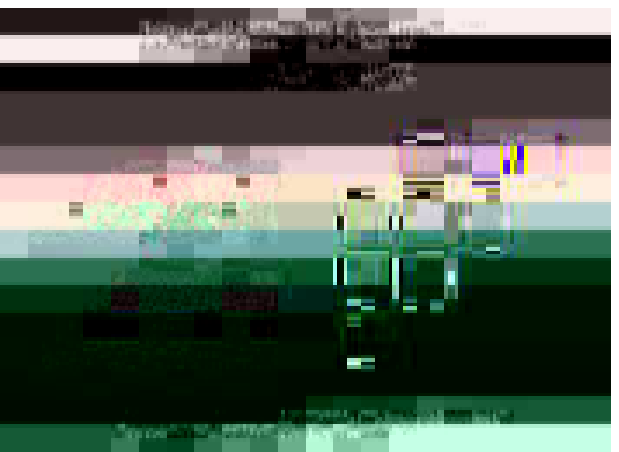}
d)
\end{minipage}
\caption{The encryption results of the 169th frame in an MPEG-1
video: a) $(p_{sr},p_{sd},p_{mv})=(0,0,0)$ -- the plain frame; b)
$(p_{sr},p_{sd},p_{mv})=(1,0,0)$; c)
$(p_{sr},p_{sd},p_{mv})=(1,1,0)$; d)
$(p_{sr},p_{sd},p_{mv})=(1,1,1)$.}\label{figure:Animation}
\end{figure}

\section{Conclusion}

This paper focuses on the problem of how to realize perceptual
encryption of MPEG videos. Based on a comprehensive survey on
related work and performance analysis of some existing perceptual
video encryption schemes, we have proposed a new design with more
useful features, such as on-the-fly encryption and multi-dimensional
perceptibility. We have also discussed its security against
deblocking attack and pointed out some measures against
known/chosen-plaintext attack. The proposed perceptual encryption
scheme can also be extended to realize non-perceptual encryption by
simply adding a VLC-encryption part.

\bibliographystyle{IEEEtran}
\bibliography{PVEA}

\end{document}